\documentclass[journal,article,submit,pdftex,moreauthors]{mdpi} 

\firstpage{1} 
\pubvolume{14}
\issuenum{1}
\articlenumber{707}
\pubyear{2024}
\copyrightyear{2024}
\datereceived{ } 
\daterevised{ } 
\dateaccepted{ } 
\datepublished{ } 
\hreflink{https://doi.org/10.3390/nano14080707} 

\Title{Pump-driven opto-magnetic properties in semiconducting transition-metal dichalcogenides: an analytical model}

\TitleCitation{Pump-driven opto-magnetic properties in semiconducting \ldots}

\Author{Habib Rostami $^{1}$\orcidA{}, Federico Cilento $^{2}$\orcidB{} and Emmanuele Cappelluti $^{3}$*\orcidC{}}

\AuthorNames{Habib Rostami, Federico Cilento and Emmanuele Cappelluti}

\AuthorCitation{Rostami, H.; Cilento, F.; Cappelluti, E.}

\address{%
$^{1}$ \quad Department of Physics, University of Bath, Claverton Down, Bath BA2 7AY, United Kingdom; hr745@bath.ac.uk\\
$^{2}$ \quad Elettra-Sincrotrone Trieste S.C.p.A., 34149 Basovizza, Italy; federico.cilento@elettra.eu\\
$^{3}$ \quad Istituto di Struttura della Materia, CNR (ISM-CNR), 34149 Trieste, Italy; emmanuele.cappelluti@ism.cnr.it}

\corres{Corresponding author}

\abstract{
Single-layer transition-metal dichalcogenides provide
an unique intrinsic entanglement between the spin/valley/orbital degrees of freedom
and the polarization of scattered photons.
This scenario gives rise to the well-assessed optical dichroism observed
by using both steady and time-resolved probes.
In this paper we provide a compact analytical modelling
of the onset of a finite Faraday/Kerr optical rotation upon shining with a circularly polarized light.
We identify different optical features displaying optical rotation at different characteristic energies,
and we describe in an analytical framework the time-dependence
of their intensities as a consequence of the main spin-conserving and spin-flip processes.
}

\begin{document}

\section{Introduction}

The family of semiconducting transition-metal dichalcogenides (TMDs) $MX_2$
($M=$Mo, W; $X=$S, Se) appears as one of the most promising
platforms for future technological applications \cite{Jariwala_2014,sun16,Manzeli_2017,arora21}.
These materials are indeed characterized by the presence of many degrees of freedom (charge, spin, valley, layer,
lattice, \ldots), strongly entangled each other \cite{mak10,sallen12,xiao12,zeng12nn,wu13,liureview15,strain15},
opening the possibility of tuning
the electronic/optical/magnetic/transport
properties in a controlled, flexible and reversible way
by external magnetic or electric fields.
When isolated at the single-layer level, these compounds
present a direct bandgap at the high-symmetry points K, K$^\prime$ of the Brillouin zone,
the {\em valleys}, as shown by photoluminescence probes \cite{mak10,splendiani10,xiao12,Schaibley_2016,Liu_2019,soni22}.
As in graphene, the honeycomb lattice structure is reflected
in peculiar optical selection rules which induce selectively
interband optical transitions in a given valley
upon circularly polarized light.
This scenario prompts the concept of ``valleytronics'', i.e. the possibility
of manipulating the quantum degrees of freedom
selectively in a single valley \cite{Schaibley_2016,Liu_2019}.
Such optical sensitivity in TMDs has been widely explored in single-layer compounds \cite{mak12nn,cao12,zeng12nn,lagarde14,plechinger14,zhu14,dalconte15,sun16,plechinger17,mccormick18,vdd18,hung19,lin21,lin22,sz20,kiemle20,arora21,caruso22}.
A common tool is the observation of an optical dichroism,
i.e. a different optical response upon left-hand or right-hand circularly polarized photons.
One striking difference of these compounds with respect to graphene
is the presence of a strong spin-orbit coupling which provides
a sizable spin-splitting of the valence band.
Within this context, circularly polarized light is not only selectively coupled
with a given valley, but also with a given spin, yielding
spin-polarized charges in the conduction band along
with opposite-spin charges in the valence band  \cite{mak12nn,cao12,zeng12nn,lagarde14,lin21,lin22,mak14science,rostami19,Wang_2013,Mai_2014,plechinger14,zhu14,dalconte15,yang15,plechinger17,molina17,mccormick18,kiemle20,arora21}.

The entanglement between light polarization and spin population
can be conveniently investigated by means of the observation of
a finite Kerr or Faraday rotation \cite{levallois12,levallois15,catarina20}.
These effects signalize the presence of an intrinsic magnetic field
in the sample, and in single-layer TMDs they can be naturally triggered
as a result of circularly polarized pumping \cite{kimel05}, which gives rise,
as discussed above, to valley-selective and spin-selective
particle-hole excitations \cite{plechinger14,zhu14,dalconte15,yang15,plechinger17,yan17,molina17,mccormick18,perlangeli20,kiemle20,arora21}.

Aim of the present paper is to provide
a compact and microscopical investigation of the onset
of Kerr/Faraday rotation in a wide energy spectrum of single-layer TMDs.
A key point is the identification of the orbital character of the
particle-hole excitations at different energies allowed by the optical selection rules,
and how it is reflected in the sign and strength of the optical Kerr/Faraday rotation.
In order to address this issue in a clearest and controlled way,
we introduce a proper generalization
of a ${\bf k} \cdot {\bf p}$ expansion in a three-band framework.
The optical response is computed at the non-interacting level
within a fully quantum Kubo approach where Kerr/Faraday effects
are related to the appearing of off-diagonal components
of the optical tensor.
Different optical features are identified as associated with different
particle-hole excitations,
and their time evolution in out-of-equilibrium dynamics are discussed.
While the exact energies of such optical features
should be considered as unrenormalized by
exciton binding effects (not considered here),
the present work provides an analytical insight
on the microscopic onset of the different optical features
whose strengths can be conveniently modelled
in terms of a unique parameter.

\section{Single-particle Hamiltonian}
\label{s:ham}

Single-layer TMDs contain a plane of $M$-atoms in a triangular lattice, sandwiched
between two layers of chalcogen atoms $X$. The resulting lattice, from a top view,
appears as a bipartite hexagonal structure.
Many theoretical approaches have been proposed to capture the relevant physics
of these materials.
As a general rule, effective low-energy models (like ${\bf k} \cdot {\bf p}$ expansions) retain only the relevant conduction
and valence bands, mapping the complex band structure onto an effective two-band gapped Dirac model \cite{xiao12,kormanyos13,kormanyos15,rostami15}.
On the other hand, tight-binding (TB) models have emphasized the role of the $d$-orbitals
of the metal atoms, in particular the $d_{3z^2-r^2}$,
$d_{xy}$, $d_{x^2-y^2}$ ones, which provide the main orbital content
of both the valence and conduction bands, as well as of a third
higher-energy conduction band \cite{rostami13,cappelluti13,zahid13,wu15,fang15,ridolfi15,dias18,jorissen24}.
From a microscopical point of view, however,
the simple triangular lattice of the $M$ atoms cannot account
for a gapped semiconducting band-structure,
and the hybridization with $X$ atoms has been shown
to play a crucial role.
The choice between a simplified, semi-analytical approach
and the multiband complexity of a fully microscopical TB model
is a delicate issue which depends on the physics on which to be focused.

An interesting balance between these two approaches has been provided 
by Liu {\em et al.} in Ref. \cite{3bands} where they introduced
a  compact three-band tight-binding model within the reduced Hilbert space:
\begin{eqnarray}
\phi_{\bf p}^\dagger
&=&
(
d_{{\bf p},3z^2-r^2}^\dagger,
d_{{\bf p},xy}^\dagger,
d_{{\bf p},x^2-y^2}^\dagger,
),
\label{phi}
\end{eqnarray}
where
the role of the hybridization of the $d$-orbitals of $M$ atoms with $p$-orbitals of $X$
is modelled, using symmetry arguments, by means of effective complex hopping parameters
that break the triangular symmetry, enforcing the physics of a bipartite hexagonal lattice.
The resulting one-particle Hamiltonian can be thus written
in the full Brillouin zone thus as:
\begin{equation}
\hat{H}_{\sigma}({\bf k})
=
\hat{H}_{{\rm TB}}({\bf k})
+
\hat{H}_{{\rm SO},\sigma}
,
\label{eq:HSO-2}
\end{equation}
where interatomic hoppings up to third-nearest-neighbor level are
included in the TB part $\hat{H}_{{\rm TB},{\bf k}}$,
providing an excellent agreement with ab-initio calculations
for the conduction and valence bands
close to the K, K$^\prime$ valleys.
The spin-orbit coupling (SOC) is safely approximated with its dominant contribution
due to the local spin-diagonal term \cite{3bands,roldan14},
which in this basis reads:
\begin{equation}
\hat{H}_{{\rm SO},\sigma}
=
\lambda I_\sigma
\begin{pmatrix}
0 & 0 & 0 \\
0 & 0 & i \\
0 & -i & 0 
\end{pmatrix}
,
\label{eq:HSO}
\end{equation}
where
$I_\sigma=\delta_{\sigma,\uparrow}-\delta_{\sigma,\downarrow}$.

Such three-band tight-binding model provides an accurate description of the
energy dispersion and of the orbital character of the main relevant bands
for optical probes with the advantage of a reduced Hilbert space.
In the following we focus on the optical response that
is governed by the particle-hole excitations close so the K, K$^\prime$ valley points.
The above model represents thus also a suitable platform for an analytical
${\bf k}\cdot {\bf p}$ expansion that generalizes the previous ${\bf k}\cdot {\bf p}$ approaches \cite{xiao12,kormanyos13,kormanyos15,rostami15}
up to the relevant three-orbital space.
To this aim we thus expand $\hat{H}({\bf k})$ up to the quadratic order
in the relative momentum ${\bf p}={\bf k}-\mbox{K}$
(${\bf p}={\bf k}-\mbox{K}^\prime$) close to the K, K$^\prime$.
It is also convenient to express the resulting Hamiltonian
$\hat{H}({\bf p},{\rm K})$, $\hat{H}({\bf p},{\rm K}^\prime)$
in the chiral basis:
\begin{eqnarray}
\psi_{\bf p}^\dagger
&=&
(
d_{{\bf p},3z^2-r^2}^\dagger,
d_{{\bf p},{\rm L}}^\dagger,
d_{{\bf p},{\rm R}}^\dagger,
),
\label{psi}
\end{eqnarray}
where $d_{{\bf p},{\rm L}}=(d_{{\bf p},x^2-y^2}-id_{{\bf p},xy})/\sqrt{2}$ and
$d_{{\bf p},{\rm R}}=(d_{{\bf p},x^2-y^2}+id_{{\bf p},xy})/\sqrt{2}$.
In such basis the spin-orbit term appears purely diagonal,
\begin{equation}
\hat{H}_{{\rm SO},\sigma}
=
-\lambda I_\sigma
\begin{pmatrix}
0 & 0 & 0 \\
0 & 1 & 0 \\
0 & 0 & -1 
\end{pmatrix}
,
\label{eq:HSOchi}
\end{equation}
and we can write:
\begin{eqnarray}
\hat{H}({\bf p},K)
&=&
\begin{pmatrix}
E_0+a_{0}p^2a^2 & v_{\rm 0/T} (p_+a) & v_{\rm 0/B} (p_-a) \\
v_{\rm 0/T} (p_+a)^* & E_{\rm T}+a_{\rm T}p^2a^2 & v_{\rm T/B} (p_+a) \\
 v_{\rm 0/B} (p_-a)^* & v_{\rm T/B} (p_+a)^* & E_{\rm B}+a_{\rm B}p^2a^2
\end{pmatrix},
\label{HpK}
\end{eqnarray}
\begin{equation}
\hat{H}({\bf p},K^\prime)
=
\begin{pmatrix}
E_0+a_{0}p^2a^2 & -v_{\rm 0/B} (p_+a) & -v_{\rm 0/T} (p_-a) \\
-v_{\rm 0/B} (p_+a)^* & E_{\rm B}+a_{\rm B}p^2a^2 & -v_{\rm T/B} (p_+a) \\
 -v_{\rm 0/T} (p_-a)^* & -v_{\rm T/B} (p_+a)^* & E_{\rm T}+a_{\rm T}p^2a^2
\end{pmatrix},
\label{HpKp}
\end{equation}
where $p_\pm=p_x\pm ip_y$, $a$ is the in-plane $M$-$M$ distance.
The full expression of the band parameters in Eqs. (\ref{HpK})-(\ref{HpKp})
in terms of the original tight-binding parameters
is provided in Appendix \ref{a:TB}.
The total spin-full Hamiltonians at the K and K$^\prime$
\begin{equation}
\hat{H}_\sigma({\bf p},\nu)
=
\hat{H}({\bf p},\nu)
+
\hat{H}_{{\rm SO},\sigma}
,
\label{hamfull}
\end{equation}
(where $\nu=$K, K$^\prime$)
contain all the relevant entanglements between spin, valleys
and chirality.

Eq. (\ref{hamfull}) defines
the energy levels at the K point
for each spin family.
We have explicitely:
\begin{eqnarray}
\epsilon_{{\rm T},\sigma}(0)
&=&
E_{\rm T}-\lambda I_\sigma
,
\\
\epsilon_{0,\sigma}(0)
&=&
E_{0}
,
\\
\epsilon_{{\rm B},\sigma}(0)
&=&
E_{\rm B}+\lambda I_\sigma
.
\end{eqnarray}

At the K point,
the energy level $E_0$ corresponds to the conduction band edge,
which results here spin-degenerate since we neglect the weak spin-orbit coupling
of the $X$ chalcogen atoms.
The
valence band at K is associated with the R-chiral state
with spin-split energies $E_{\rm B}\pm\lambda$,
for up and down spin, respectively.
The $E_{\rm T}\pm\lambda$ levels correspond to higher energy states,
characterized by a L-chiral symmetry \cite{cappelluti13}.
A similar energy spectrum is found at the K$^\prime$ point,
but with chiral content exchanged between the valence and the high-energy levels.

The optical selection rules are encoded in the multiband matrix structure
of the current operators
which can be straightforwardly computed
as derivatives of the single-particle Hamiltonian, $\hat{J}_i({\bf p},\nu)=dH({\bf p},\nu)/dp_i$,
where $i=x,y$ and $\nu=$K, K$^\prime$.
At the high-symmetry points (${\bf p}=0$) we get:
\begin{eqnarray}
\hat{J}_x(K)
&=&
\begin{pmatrix}
0 & v_{\rm 0/T} a & v_{\rm 0/B} a\\
v_{\rm 0/T}a & 0 & v_{\rm T/B}a \\
 v_{\rm 0/B}a & v_{\rm T/B}a & 0
\end{pmatrix},
\label{jxk}
\\
\hat{J}_y(K)
&=&
\begin{pmatrix}
0 & iv_{\rm 0/T} a & -iv_{\rm 0/B} a\\
-iv_{\rm 0/T} a& 0 & iv_{\rm T/B}a \\
 iv_{\rm 0/B}a & -iv_{\rm T/B} a& 0
\end{pmatrix}
,
\label{jyk}
\\
\hat{J}_x(K^\prime)
&=&
-
\begin{pmatrix}
0 & v_{\rm 0/B} a & v_{\rm 0/T} a\\
v_{\rm 0/B}a & 0 & v_{\rm T/B}a \\
 v_{\rm 0/T}a & v_{\rm T/B}a & 0
\end{pmatrix}
,
\\
\hat{J}_y(K^\prime)
&=&
\begin{pmatrix}
0 & -iv_{\rm 0/B} a & iv_{\rm 0/T} a\\
iv_{\rm 0/B} a& 0 & -iv_{\rm T/B}a \\
 -iv_{\rm 0/T}a & iv_{\rm T/B} a& 0
\end{pmatrix}
.
\end{eqnarray}

In similar way, one can derive the chiral current operators
$\hat{J}_\pm(\nu)=\hat{J}_x(\nu)\pm i \hat{J}_x(\nu)$:
\begin{eqnarray}
\hat{J}_+(K)
&=&
\begin{pmatrix}
0 & 0 & 2v_{\rm 0/B} a  \\
2v_{\rm 0/T}a & 0 & 0 \\
0 & 2v_{\rm T/B} & 0
\end{pmatrix}
,
\label{jpK}
\\
\hat{J}_-(K)
&=&
\begin{pmatrix}
0 & 2v_{\rm 0/T} a & 0 \\
0 & 0 & 2v_{\rm T/B} \\
 v_{\rm 0/B}a & 0 & 0
\end{pmatrix}
,
\\
\hat{J}_+(K')
&=&
\begin{pmatrix}
0 & 0 & -2v_{\rm 0/T} a\\
-v_{\rm 0/B}a & 0 & 0 \\
 0 & -v_{\rm T/B}a & 0
\end{pmatrix}
,
\\
\hat{J}_-(K')
&=&
\begin{pmatrix}
0 & -2v_{\rm 0/B} a & 0\\
0 a& 0 & -2v_{\rm T/B}a \\
 -2v_{\rm 0/T}a & 0 a& 0
\end{pmatrix}
.
\label{jmKp}
\end{eqnarray}

In order to evaluate the optical response,
Eq. (\ref{hamfull}) can be numerically diagonalized for finite momentum ${\bf p}$
to obtain eigenvalues (the band dispersion)
and eigenvectors of the resulting eigenstates.
This task can be however further simplified within a ${\bf k}\cdot {\bf p}$ framework
where the band dispersion, at the quadratic order we are interested in,
is simply provided by the diagonal terms of Eqs. (\ref{HpK})-(\ref{HpKp}) {\em corrected}
by the second-order corrections resulting from the off-diagonal elements.
We get thus:
\begin{eqnarray}
\hat{H}_\sigma({\bf p},\nu)
&\approx&
\hat{E}_\sigma({\bf p},\nu)
,
\end{eqnarray}
where
\begin{eqnarray}
\hat{E}_\sigma({\bf p},K)
&=&
\left(
\begin{array}{ccc}
\epsilon_{0,\sigma}(p) & 0 & 0 \\
0 & \epsilon_{{\rm T},\sigma}(p) & 0 \\
 0 & 0 & \epsilon_{{\rm B},\sigma}(p)
\end{array}
\right),
\end{eqnarray}
\begin{eqnarray}
\hat{E}_\sigma({\bf p},K^\prime)
&=&
\left(
\begin{array}{ccc}
\epsilon_{0,-\sigma}(p) & 0 & 0 \\
0 & \epsilon_{{\rm B},-\sigma} & 0 \\
 0 & 0 & \epsilon_{{\rm T},-\sigma}(p)
\end{array}
\right),
\end{eqnarray}
and where 
\begin{eqnarray}
\epsilon_{T,\sigma}(p)
&=&
E_{\rm T}-\lambda I_\sigma
+
\bar{a}_{{\rm T},\sigma}p^2a^2
,
\label{ets}
\\
\epsilon_{0,\sigma}(p)
&=&
E_{0}
+
\bar{a}_{0,\sigma}p^2a^2
,
\\
\epsilon_{{\rm B},\sigma}(p)
&=&
E_{\rm B}+\lambda I_\sigma
+\bar{a}_{{\rm B},\sigma}p^2a^2
.
\label{ebs}
\end{eqnarray}
The numerical expression of the ${\bf k}\cdot {\bf p}$
band parameters in Eqs. (\ref{ets})-(\ref{ebs}) is also provided
in Appendix \ref{a:TB}.
\begin{figure}[t]
\begin{center}
\includegraphics[width=12.5 cm]{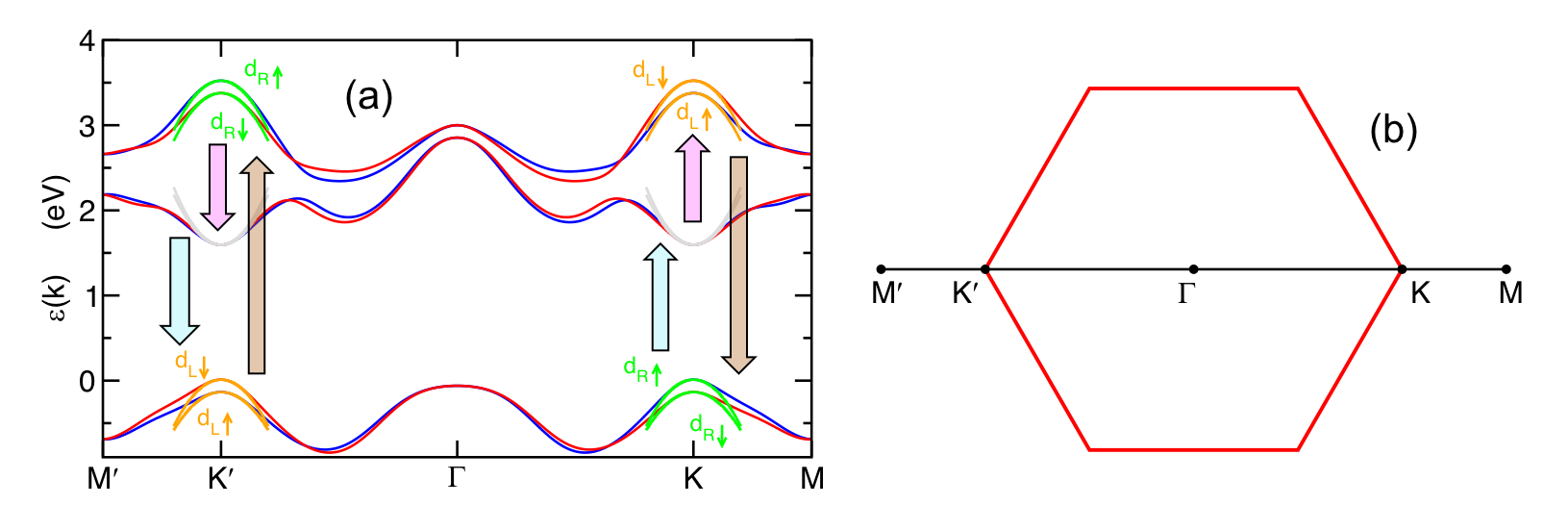}
\caption{(\textbf{a}) Comparison between the three-band TB model from Ref. \cite{3bands}
and our analytical three-band ${\bf k} \cdot {\bf p}$ model for single-layer MoS$_2$ along the path described
in panel (\textbf{b}).
Blue and red solid lines represent the TB band-dispersion for spin-up and spin-down electrons,
respectively. Green, orange, and grey lines show the band-dispersions
for the eigenstates with orbital character $d_{\rm R}$, $d_{\rm L}$, and $d_{3z^2-r^2}$, 
respectively.
The vertical arrows represent the optical interband transitions allowed
at the K, K$^\prime$ points by a left-circularly polarized photon.
\label{fl-disp}
}
\end{center}
\end{figure}   
\unskip
The comparison between the full TB dispersion in the Brillouin zone, from Ref. \cite{3bands},
and our analytical three-band model expanded around
the K, K$^\prime$ points is displayed in Fig. \ref{fl-disp}, showing
an excellent agreement.

Note that, within the same  ${\bf k}\cdot {\bf p}$ expansion,
the leading order to the current operators is not affected
and Eqs. (\ref{jpK})-(\ref{jmKp}) are still valid
also in the ${\bf k}\cdot {\bf p}$ context.
As mentioned, the different matrix structure of Eqs. (\ref{jpK})-(\ref{jmKp})
enforces the different optical selection rules at the K and K$^\prime$ points.
It is useful to recall that right-circularly polarized light (RCP) couples with the L-chiral current $J_-$,
whereas LCP couples with $J_+$, according with the relation: $A_x J_x+A_y J_y= [A_+ J_-+A_- J_+]/2$.
Eqs. (\ref{jpK})-(\ref{jmKp}) dictate for instance how, under external pumping conditions,
absorption of a left-circularly polarized (LCP) photon
can induce particle-hole optical excitations
at the K point only between the valence band with $d_{\rm R}$ character
and the conduction band with $d_{3z^2-r^2}$ character,
or (in case of electron-doped samples) between the conduction band with $d_{3z^2-r^2}$ character
and the high-level conduction band with $d_{\rm L}$ character.
On the other hand, the same absorption of a LCP photon can
effectively create particle-hole optical excitations
at the K$^\prime$ point between the valence band with $d_{\rm R}$ character
and the high-level conduction band with $d_{\rm L}$ character.
The selection rules for right-circularly polarized light, coupled
with the chiral current $J_+$, are graphically obtained
by reversing the arrow of each particle-hole excitations.

Similar selection rules govern also the virtual processes involved
in the optical linear response, whose analytical computation will be addressed
in the next section.

\section{Optical response}

Equipped with the theoretical tools outlines in Section \ref{s:ham}, we can now compute in the useful chiral basis all the elements
of the optical tensor of single-layer transition-metal dichalcogenides through the evaluation of the frequency-dependent current-current response function.
In the Matsubara space we have:
\begin{eqnarray}
\pi_{ij}(i\hbar\omega_m,\nu)
&=&
 \frac{e^2 T}{4\pi^2 \hbar^2} 
 \sum_{\sigma,n}
\int d^2{\bf p}
\mbox{Tr}
\left[
\hat{J}_i(\nu)
\hat{G}_\sigma({\bf p},i\omega_n+i\omega_m,\nu)
\hat{J}_j^\dagger(K)
\hat{G}_\sigma({\bf p},i\omega_n,\nu)
\right],
\end{eqnarray}
where $i,j=x,y$, $\nu=$K, K$^\prime$,
\begin{eqnarray}
\hat{G}_\sigma({\bf p},z,\nu)
&=&
\frac{1}{(z+\mu)\hat{I}-\hat{E}_\sigma({\bf p},\nu)}
,
\end{eqnarray}
and where $\mu$ is the chemical potential.
Here $\hbar \omega_n=\pi T(2n+1)$ are the internal fermionic frequencies,
$\hbar \omega_m=2\pi T m$ is the external bosonic frequency,
and $T$ is the temperature.

The generic elements of the optical conductivity tensor are thus obtained as:
\begin{eqnarray}
\sigma_{ij}(\omega,\nu)
&=&
-\frac{\pi_{ij}(\hbar\omega,\nu)}{i\omega}
,
\end{eqnarray}
where
\begin{eqnarray}
\pi_{ij}(\hbar\omega,\nu)
&=&
\pi_{ij}(i\hbar \omega_m,\nu)|_{i\omega_m \to \omega}
.
\end{eqnarray}

Given the three-band structure of our model, the total optical response can be divided
(leaving aside the Drude-like intraband terms at low frequencies which are not relevant
in the present analysis) in three interband contributions:
\begin{eqnarray}
\sigma_{ij}(\omega,\nu)
&=&
\sigma_{ij}^{\rm 0,T}(\omega,\nu)
+
\sigma_{ij}^{\rm 0,B}(\omega,\nu)
+
\sigma_{ij}^{\rm T,B}(\omega,\nu)
.
\end{eqnarray}
The term $\sigma_{ij}^{\rm 0,B}$ describes optical transitions
between the (spin-split) valence band and the lowest energy conduction bands.
Due to the spin-splitting of the valence band, this term accounts
for the A and B exciton resonances.
The term $\sigma_{ij}^{\rm T,B}$ conveys information
about optical transitions
between the valence band (with $d_{\rm R}$ or $d_{\rm L}$ character) and the high-energy conduction band
with opposite $d_{\rm R}/d_{\rm L}$ character, which has been discussed in details in Refs. \cite{lin21,lin22}.
Finally, the term $\sigma_{ij}^{\rm 0,T}$ describes
optical transitions between the lowest-energy and high-energy conductions bands.
In the undoped semiconducting regime,
due to the Pauli blocking,
this term is usually irrelevant, but it plays a role upon photo-induced doping \cite{lin21,lin22}.

All the terms present a similar functional structure 
where the relevant role is played by the band population.
For sake of simplicity, we focus thus for the moment on the first term $\sigma_{ij}^{\rm 0,B}$.
Since the system is block-diagonal in the spin-index, one can also
formally compute separately the optical response $\sigma_{ij,\sigma}(\omega,\nu)$ for each spin and each valley $\nu$.
We have for instance:
\begin{eqnarray}
\sigma_{xx,\sigma}^{\rm 0,B}(\omega,{\rm K})
&\approx&
-
\frac{e^2}{4\pi^2 \hbar^2}
\frac{(v_{\rm 0/B} a)^2}{\omega}
\int d^2{\bf p}
\Bigg[
\frac{
f[\epsilon_{0,\sigma}(p)]-f[\epsilon_{{\rm B},\sigma}(p)]
}{\epsilon_{0,\sigma}(p)-\epsilon_{{\rm B},\sigma}(p)-\omega-i\delta}
\nonumber\\
&&
+
\frac{
f[\epsilon_{0,\sigma}(p)]-f[\epsilon_{{\rm B},\sigma}(p)]
}{\epsilon_{0,\sigma}(p)-\epsilon_{{\rm B},\sigma}(p)+\omega+i\delta}
\Bigg]
,
\label{sigmaxx}
\end{eqnarray}
where $f[E]$ is the occupation factor for given momentum and band,
which under equilibrium conditions, $f[E]=1/\{{\rm exp}[(E-\mu)/T]+1\}$.
At $T=0$, in the semiconducting state $\mu=0$,
one get $f[\epsilon_{0,\sigma}(p)]=0$, $f[\epsilon_{{\rm B},\sigma}(p)]=1$ and,
for $\omega>0$:
\begin{eqnarray}
\mbox{Re}\sigma_{xx,\uparrow}^{\rm 0,B}(\omega,{\rm K})
&=&
\sigma_0
\frac{ v_{\rm 0/B}^2}{c_{\rm A}\hbar\omega}
\theta(\hbar \omega-\Delta_{\rm A})
,
\\
\mbox{Re}\sigma_{xx,\downarrow}^{\rm 0,B}(\omega,{\rm K})
&=&
\sigma_0
\frac{v_{\rm 0/B}^2}{c_{\rm B}\hbar \omega}
\theta(\hbar \omega-\Delta_{\rm B})
,
\end{eqnarray}
where $\sigma_0=e^2/4\hbar$ is the universal two-dimensional conductivity,
$\Delta_{\rm A}$, $\Delta_{\rm B}$ and the excitation edges for the A and B excitons,
respectively,
\begin{eqnarray}
\Delta_{\rm A}
&=&
E_0-E_{\rm B} -\lambda
,
\\
\Delta_{\rm B}
&=&
E_0-E_{\rm B} +\lambda
,
\end{eqnarray}
and
\begin{eqnarray}
c_{\rm A}
&=&
\bar{a}_{0,\uparrow}
-
\bar{a}_{\rm B,\uparrow}
,
\end{eqnarray}
\begin{eqnarray}
c_{\rm B}
&=&
\bar{a}_{0,\downarrow}
-
\bar{a}_{\rm B,\downarrow}
.
\end{eqnarray}
Using the tight-binding parameters of Ref. \cite{3bands},
we get for MoS$_2$ $\Delta_{\rm A}=1.584$ eV, $\Delta_{\rm B}=1.730$ eV, 
$v_{\rm 0/B}=$,
$c_{\rm A}=1.598$ eV, and $c_{\rm B}=1.616$ eV.
For symmetry we have also $\sigma_{xx,\sigma}(\omega,\nu)=\sigma_{yy,\sigma}(\omega,\nu)$.

In similar way, one gets:
\begin{eqnarray}
\sigma_{xy,\sigma}^{0,\rm B}(\omega,K)
&=&
i
\frac{e^2}{4\pi^2 \hbar^2}
\frac{(v_{\rm 0/B} a)^2}{\omega}
\sum_{\bf p}
\Bigg[
\frac{
f[\epsilon_{0,\sigma}(p)]-f[\epsilon_{{\rm B},\sigma}(p)]
}{\epsilon_{0,\sigma}(p)-\epsilon_{{\rm B},\sigma}(p)-\omega-i\delta}
\nonumber\\
&&
-
\frac{
f[\epsilon_{0,\sigma}(p)]-f[\epsilon_{{\rm B},\sigma}(p)]
}{\epsilon_{0,\sigma}(p)-\epsilon_{{\rm B},\sigma}(p)+\omega+i\delta}
\Bigg]
,
\label{sigmaxy}
\end{eqnarray}
and, for $T,\mu=0$ and $\omega>0$:
\begin{eqnarray}
\mbox{Im}\sigma_{xy,\uparrow}^{0,\rm B}(\omega,{\rm K})
&=&
-
\sigma_0
\frac{v_{\rm 0/B}^2}{c_{\rm A}\hbar\omega}
\theta(\hbar\omega-\Delta_{\rm A})
,
\\
\mbox{Im}\sigma_{xy,\downarrow}^{0,\rm B}(\omega,{\rm K})
&=&
-
\sigma_0
\frac{v_{\rm 0/B}^2}{c_{\rm B} \hbar \omega}
\theta(\hbar\omega-\Delta_{\rm B})
.
\end{eqnarray}

\begin{figure}[t!]
\begin{center}
\includegraphics[width=0.95\textwidth]{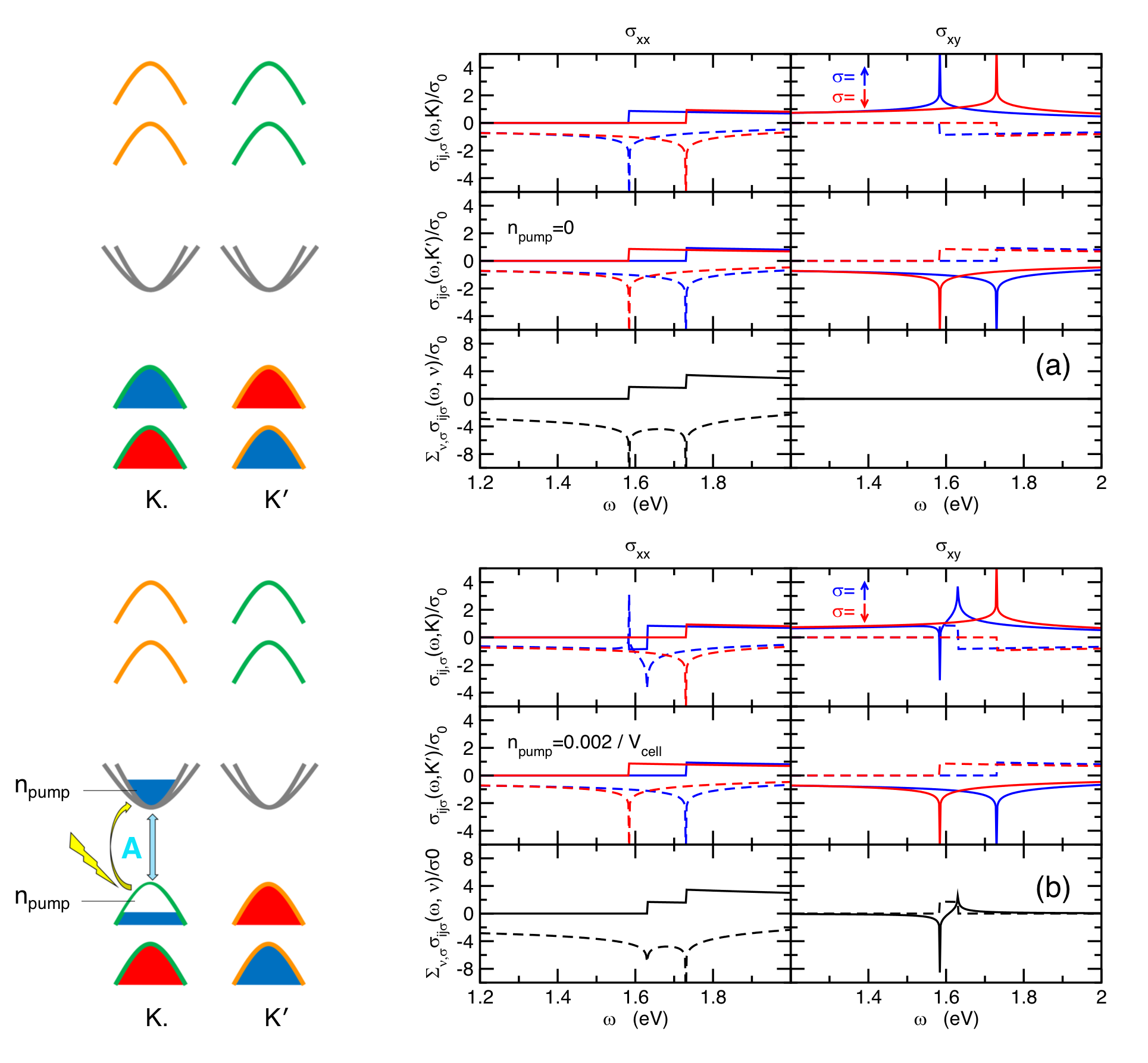}
\end{center}
\caption{Plot of the different contributions to the diagonal $\sigma_{xx}$
and off-diagonal part $\sigma_{xy}$ of the optical conductivity of single-layer TMDs (e.g. MoS$_2$ here)
resolved for different valleys
and different spins. Panel ({\bf a}) represents the semiconducting equilibrium case, panel ({\bf b})
the non-equilibrium case induced by a left-circularly polarized pumping tuned at the $\omega=\Delta_{\rm A}$ frequency,
which acts only on the top valence band at the K point. Blue and red lines represent contributions from spin up and down,
respectively, whereas black lines the total optical conductivity summed up over spin and valleys.}
\label{f-panels}
\end{figure}

The real parts of $\sigma_{xy,\sigma}(\omega,{\rm K})$
and the imaginary parts of $\sigma_{xx,\sigma}(\omega,{\rm K})$
can be thus easily obtained using the Kramers-Kronig relations.
Furthermore, using the symmetry relations encoded in the different matrix structures at different valleys,
one can recognize that at equilibrium:
\begin{eqnarray}
\sigma_{xx,\sigma}^{0,\rm B}(\omega,{\rm K}^\prime)
&=&
\sigma_{xx,-\sigma}^{0,\rm B}(\omega,{\rm K})
,
\end{eqnarray}
\begin{eqnarray}
\sigma_{xy,\sigma}^{0,\rm B}(\omega,{\rm K}^\prime)
&=&
-
\sigma_{xy,-\sigma}^{0,\rm B}(\omega,{\rm K})
,
\end{eqnarray}
($-\sigma$ being here the reversed $\sigma$ spin),
so that, under such equilibrium conditions, the contributions of different valleys {\em sum up}
for the diagonal terms of the optical tensor,
whereas they {\em cancel out} for the off-diagonal ones,
in accordance with the time-reversal symmetry.
The net result for the whole optical tensor
is summarized in Fig. \ref{f-panels}a.
Similar expressions can be derived for 
the $\sigma_{xy,\sigma}^{0,\rm T}(\omega,\nu)$
and
$\sigma_{xy,\sigma}^{\rm T,\rm B}(\omega,\nu)$
terms.

\section{Non-equilibrium optical response}

Eqs. (\ref{sigmaxx}), (\ref{sigmaxy}) provide a suitable framework to model the optical response
under non equilibrum conditions, by specifying the proper occupation factors
in the presence of photo-induced particle-hole excitations.
More specifically, within such semi-classical approach,
we can simulate the effects of a LCP laser pumping tuned at the $\omega \approx \Delta_{\rm A}$ frequency
by assuming an effective photoinduced charge transfer $n_{\rm pump}$ from the valence 
to the conduction band. Due to the selected circular polarization, such particle-hole
excitations occur only at the K point and, due to the selected frequency in resonance with the A exciton,
only for the spin $\uparrow$.
Since only one valley, with a given spin, will be populated in both the valence and conduction band,
the photoinduced charge density can be further parametrized in terms of a characteristic momentum $\bar{p}$,
for which $f[\epsilon_{0,\sigma}(p)]=1$, $f[\epsilon_{{\rm B},\sigma}(p)]=0$ ($p \le \bar{p}$),
obeying the relation:
\begin{eqnarray}
n_{\rm pump}
&=&
\frac{\bar{p}^2}{4\pi}
.
\end{eqnarray}
Typical values of $n_{\rm pump}$ can range up to $n_{\rm pump} \lesssim 0.01 /  V_{\rm cell}$.
For a representative case $n_{\rm pump} = 0.002 /  V_{\rm cell}$,
we get for MoS$_2$ $\bar{p} a \approx 0.171$,
and, using $a=3.16$ \AA, $\bar{p}\approx 0.054$ \AA$^{-1}$.

Due to the selection rules, only the states at K point with the proper chirality,
corresponding in this case to spin up, are affected by the pumping.
For $|p| \le \bar{p}$ we have thus a {\em reverse} Pauli blocking.
The contribution of these states to the optical conductivity reads thus:
\begin{eqnarray}
\mbox{Re}\sigma^{\rm 0/B}_{xx,\uparrow,{\rm pump}}(\omega,{\rm K})
&=&
\sigma_0
\frac{v_{\rm 0/B}^2}{c_{\rm A}\hbar\omega}
\left[
-
\theta(\hbar\omega-\Delta_{\rm A})
+
2\theta(\hbar\omega-\Delta_{\rm A}-E_{\rm P,A})
\right]
,
\\
\mbox{Im}\sigma^{\rm 0/B}_{xy,\uparrow,{\rm pump}}(\omega,{\rm K})
&=&
-
\sigma_0
\frac{v_{\rm 0/B}^2}{c_{\rm A}\hbar\omega}
\left[
-
\theta(\hbar\omega-\Delta_{\rm A})
+
2\theta(\hbar\omega-\Delta_{\rm A}-E_{\rm P,A})
\right]
,
\end{eqnarray}
where $E_{\rm P,A}=c_{\rm A} (\bar{p} a)^2=4\pi c_{\rm A}a^2n_{\rm pump}$.
In Fig. \ref{f-panels}b we show the effect
of the left-circularly polarized pumping on the diagonal
and off-diagonal parts of the optical conductivity,
$\sigma_{ij,{\rm pump}}(\omega)
=
\sum_{\sigma,\nu}
\sigma^\prime_{xx,\sigma,{\rm pump}}(\omega,\nu)$.
Due to the reverse Pauli blocking,
the real part of the diagonal term Re$\sigma_{xx,{\rm pump}}(\omega)$
shows a depletion of spectral intensity close to the A-edge energy.
In real samples, in the presence of many-body exciton binding,
this depletion appears as a reduction of the A-exciton intensity,
as been experimentally observed many times in reflectivity probes.
More striking is the result on the off-diagonal component of the optical tensor
$\sigma_{xy,{\rm pump}}(\omega)$ where
the contributions
from spin-up and spin-down transitions and from K and K$^\prime$ close to
the A-resonance do not cancel out
anymore, giving rise to a {\em finite} off-diagonal term $\sigma_{xy,{\rm pump}}(\omega) \neq 0$,
\begin{eqnarray}
\mbox{Im}\sigma^{\rm 0/B}_{xy,{\rm pump}}(\omega)
&=&
\sigma_0
\frac{2 v_{\rm 0/B}^2}{c_{\rm A}\hbar\omega}
\left[
\theta(\hbar\omega-\Delta_{\rm A})
-
\theta(\hbar\omega-\Delta_{\rm A}-E_{\rm P,A})
\right].
\label{exy}
\end{eqnarray}

On the experimental ground, the onset of a finite
off-diagonal component $\sigma_{xy,{\rm pump}}(\omega) \neq 0$
is observed as a optical (Faraday or Kerr) rotation of the transmitted/reflected polarization
of the probe, which commonly signalizes the presence of a finite effective magnetic field \cite{levallois12,levallois15,catarina20}.
More in details, we estimate a off-diagonal spectral intensity at the A-exciton energy range:
\begin{eqnarray}
I_{\rm A}
&=&
\int_{\Omega_{\rm A}} d\omega \,
\mbox{Im}\sigma_{xy,{\rm pump}}^{\rm 0/B}(\omega)
=
\sigma_0
\frac{v_{\rm 0/B}^2}{\hbar c_{\rm A}\Delta_{\rm A}}
2E_{\rm P,A}
=
4\pi a^2 
\sigma_0
\frac{v_{\rm 0/B}^2}{\hbar \Delta_{\rm A}}
2 n_{\rm pump}
,
\label{IkerrA}
\end{eqnarray}
where the spectral off-diagonal intensity is meant
to be integrated in a frequency range $\Omega_{\rm A}$ close to the A-resonance.
We stress here that, although we employ here a non-interacting model to get an analytical insight,
the physical origin of such magneto-optical rotation depends uniquely on
the selective valley-population enforced by the circularly-polarized pumping,
yielding a non-equivalent optical response 
that does not cancel in the K and K$^\prime$ valleys.
This is a quite general mechanism that will not be affected by
the formation of localized states when bound excitons form.
Within this context, we expect that the step-function spectral shape
of Eq. (\ref{exy}), also shown in Fig. \ref{f-panels}b,
will evolve smoothly in a $\delta$-like Lorenzian peak,
preserving an integrated intensity $I_{\rm A}$ which
is dictated by the amount of the spin-polarized photo-induced charges
in the conduction and valence bands, and hence still scaling with $n_{\rm pump}$.
The onset of a Faraday/Kerr rotation at the A-exciton energy
is consistent with previous experimental and theoretical investigations
\cite{plechinger14,zhu14,dalconte15,sun16,plechinger17,mccormick18,catarina20}.
It is worth underlining here that the valley-selective/spin-selective population
induced by the circularly-polarized pumping is expected within our modelling to give rise to a finite pump-driven
Kerr/Faraday rotation also at two further energy scales, which we identify with the so-called C-exciton
and with another characteristic energy which we denote as D-peak.

We notice a {\em remarkably strong} band-nesting between these two bands close to the K/K$^\prime$ points.
Such feature has been usually disregarded in the context of TMDs, where the analyses
have focused on the nesting properties between the.valence and the lowest-energy conduction
band \cite{carvalho13,gibertini14,kozawa14}.
We relate these transitions with the broad shoulder commonly denoted as C-exciton.

Currently, the origin of the remarkable shoulder in the optical conductivity
denoted as C-exciton has not been fully assessed.
A mainstream consensus associates this spectral feature with the enhanced optical activity
between the valence band and the lowest-energy conduction band along the $\Gamma$-K path,
where these two bands are thought to have a parallel energy dispersion ({\em band nesting}) \cite{carvalho13,gibertini14,kozawa14}.
Generalizing this idea within the three-band context,
we notice a {\em remarkably strong} band-nesting at K/K$^\prime$ points
between the valence and the lowest-energy conduction, governed by the
nesting factor $\sim 1/|\omega-\epsilon_{{\rm T},\sigma}(p)+\epsilon_{{\rm B},\sigma}(p) |$.
Prompted by a careful analysis
of the first-principle and tight-binding dispersions, we suggest thus a slightly different
perspective, where the C-exciton shoulder stems from band-nesting
close to the K (K$^\prime$) point between the valence band with $d_{\rm R}$ ($d_{\rm L}$) character
and the high-energy conduction band with $d_{\rm L}$ ($d_{\rm R}$) character.
In our modelling, neglecting the exciton binding energy,
we can expect thus $\Delta_{\rm C}=E_{\rm T}-E_{\rm B} -2\lambda$.
Such change of perspective has a deep impact on the predictions about
the effects of pumping with circularly-polarized light
on the  magneto-optical properties.
In the original scenario, the ${\bf k}$ points responsible for the band-nesting
are located close to the $\Gamma$ point along the path $\Gamma$-K.
These states do not have a significant chiral character,
as a consequence they have a small spin-splitting and they
are weakly entangled with circularly-polarized light.
On the contrary, in the present context where band-nesting states
lie close to the K, K$^\prime$ points, we can predict a strong chiral character,
a different response for spin-up and spin-down charges,
a strong entanglement with the circularly-polarized light
and a remarkable onset of a off-diagonal component of the optical tensor.
Such picture is consistent with the experimental findings observed in Refs. \cite{lin21,lin22}.

\begin{figure}[t!]
\begin{center}
\includegraphics[width=0.95\textwidth]{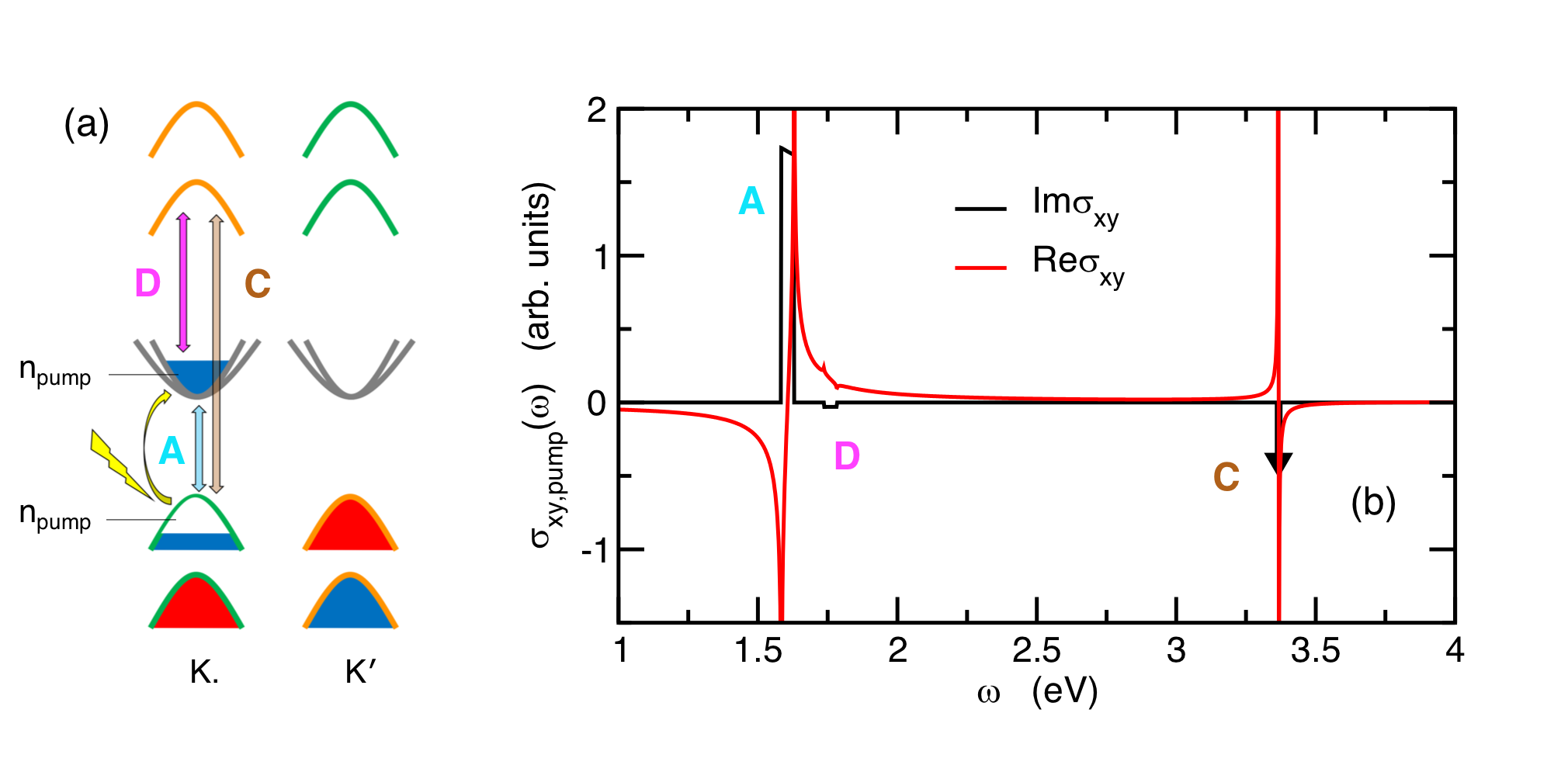}
\end{center}
\caption{({\bf a}) Sketch of the optical processes responsible for the Faraday/Kerr rotation
at different energies driven by the circularly-polarized pumping, which is parametrized
in terms of a charge density $n_{\rm pump}$ tranferred from the valence to the conduction band
at the K point. The vertical light-blue arrow marks the transitions associated with the
A-exciton from the valence band with orbital character
$d_{\rm R}$ to the lowest conduction band with $d_{3z^2-r^2}$ character.
The brown arrow marks the transitions associated with the
C-exciton from the valence band with orbital character
$d_{\rm R}$ to the high-energy conduction band with $d_{\rm L}$ character.
These processes profit of the strong band-nesting between these two bands
close to the K, K$^\prime$ points.
The magenta arrow marks the transitions giving rise
to an additional optical feature, denoted as D-peak,
corresponding to the transitions between
lowest conduction band with $d_{3z^2-r^2}$ character
and the high-energy $d_{\rm L}$-band.
({\bf b}) Real and imaginary part of the off-diagonal component
of the optical tensor $\sigma_{xy,{\rm pump}}(\omega)$
as driven by the circularly-polarized pumping.
Due to the strong band-nesting,
the C-exciton feature in Im$\sigma_{xy,{\rm pump}}$
is too narrow and sharp to be visible on this scale, and it has been represented
by the thick black arrow.
}
\label{f-CD}
\end{figure}

Our three-band model is naturally suited to describe this scenario,
where the band-nesting optical transitions responsible for the C-exciton shoulder
are accounted by the $\sigma_{ij,{\rm pump}}^{\rm T/B}(\omega)$ term
(see Fig. \ref{f-CD}a).
At the same time, the photo-induced charging of the conduction band
triggers finite optical transitions between the conduction band itself with $d_{3z^2-r^2}$
and the high-energy conduction band with $d_{\rm L}$ ($d_{\rm R}$) character.
The valley-population of these states is also very sensitive to circularly-polarized light,
and they are expected thus to drive a finite optical rotation at
typical energy, neglecting the exciton binding energy,
$\Delta_{\rm D}=E_{\rm T}-E_0 -\lambda
=\Delta_{\rm C}-\Delta_{\rm A}$ (Fig. \ref{f-CD}a).
These latter optical transitions are taken into account
by the term $\sigma_{ij,{\rm pump}}^{\rm 0/T}(\omega)$.

The effect of photo-induced pump charging
with circularly polarized light in the whole frequency domain
can be thus evaluated by considering all the possible contributions,
$\sigma_{ij,{\rm pump}}(\omega)=\sigma_{ij,{\rm pump}}^{\rm 0/B}(\omega)+\sigma_{ij,{\rm pump}}^{\rm T/B}(\omega)+\sigma_{ij,{\rm pump}}^{\rm 0/T}(\omega)$.
The formal structure of each term is very similar to the
term $\sigma_{ij,{\rm pump}}^{\rm 0/B}(\omega)$ which we have explicitly evaluated above.
Taking into account the slight differences for each term, we get thus:
\begin{eqnarray}
\mbox{Im}\sigma_{xy,{\rm pump}}(\omega)
&=&
\sigma_0
\frac{2v_{\rm 0/B}^2}{c_{\rm A}\hbar\omega}
\left[
\theta(\hbar\omega-\Delta_{\rm A})
-
\theta(\hbar\omega-\Delta_{\rm A}-E_{\rm P,A})
\right]
\nonumber\\
&&
-
\sigma_0
\frac{v_{\rm T/B}^2}{\left|c_{\rm C}\right|\hbar\omega}
\left[
\theta(\hbar\omega-\Delta_{\rm C})
-
\theta(\hbar\omega-\Delta_{\rm C}-E_{\rm P,C})
\right]
\nonumber\\
&&
-
\sigma_0
\frac{v_{\rm 0/T}^2}{\left|c_{\rm D}\right|\hbar\omega}
\left[
\theta(\hbar\omega-\Delta_{\rm }+E_{\rm P,D})
-
\theta(\hbar\omega-\Delta_{\rm D})
\right]
,
\label{exyfull}
\end{eqnarray}
where
\begin{eqnarray}
c_{\rm C}
&=&
\bar{a}_{{\rm T},\uparrow}
-
\bar{a}_{{\rm B},\uparrow}
,
\end{eqnarray}
\begin{eqnarray}
c_{\rm D}
&=&
\bar{a}_{{\rm T},\uparrow}
-
\bar{a}_{0,\uparrow}
,
\end{eqnarray}
and where 
$E_{\rm P,C}=|c_{\rm C}| (\bar{p} a)^2=4\pi |c_{\rm C}|a^2n_{\rm pump}$,
$E_{\rm P,D}=|c_{\rm D}| (\bar{p} a)^2=4\pi |c_{\rm D}|a^2n_{\rm pump}$.
The real part $\mbox{Re}\sigma_{xy,{\rm pump}}(\omega)$
is thus obtained by Kramers-Kronig relations.
In Eq. (\ref{exyfull}) we have assumed that $c_{\rm C}>0$,
which is a valid assumption for the W-based transition-metal
dichalcogenides WS$_2$, WSe$_2$, WTe$_2$ (see Table
\ref{t:par}).
However, since these materials are very close to the perfect
band-nesting limit ($c_{\rm C} \approx 0$)
for these states at the K, K$^\prime$ point,
the sign of $c_{\rm C}$ is not {\em a priori} determined.
For the Mo$X_2$ family for instance $c_{\rm C}<0$
(see Table \ref{t:par}), and the analytical expression
$\mbox{Im}\sigma_{xy,{\rm pump}}(\omega)$
should rather read:
\begin{eqnarray}
\mbox{Im}\sigma_{xy,{\rm pump}}(\omega)
&=&
\sigma_0
\frac{2 v_{\rm 0/B}^2}{c_{\rm A}\hbar \omega}
\left[
\theta(\hbar\omega-\Delta_{\rm A})
-
\theta(\hbar\omega-\Delta_{\rm A}-E_{\rm P,A})
\right]
\nonumber\\
&&
-
\sigma_0
\frac{v_{\rm T/B}^2}{\left|c_{\rm C}\right|\hbar\omega}
\left[
\theta(\hbar\omega-\Delta_{\rm C}+E_{\rm P,C})
-
\theta(\hbar\omega-\Delta_{\rm C})
\right]
\nonumber\\
&&
-
\sigma_0
\frac{v_{\rm 0/T}^2}{\left|c_{\rm D}\right|\hbar\omega}
\left[
\theta(\hbar\omega-\Delta_{\rm }+E_{\rm P,D})
-
\theta(\hbar\omega-\Delta_{\rm D})
\right]
.
\label{exyfull2}
\end{eqnarray}
The plot of $\sigma_{xy,{\rm pump}}(\omega)$ for MoS$_2$,
with a pump-driven photo-induced charge density
$n_{\rm pump} = 0.002 /  V_{\rm cell}$
($\bar{p} a \approx 0.171$)
is shown in Fig. \ref{f-CD}b,
showing how a valley-selective population due to
a circularly-polarized pumping gives rise to a finite off-diagonal component
(and hence to a finite Farady/Kerr effect) not only at the A-exciton energy $\Delta_{\rm A}$,
but also at the C-exciton energy $\Delta_{\rm C}$ and at another energy range $\Delta_{\rm D}$
corresponding to $\Delta_{\rm D}\approx \Delta_{\rm C}-\Delta_{\rm A}$, net of the exciton binding energy.
We predict thus an opposite sign of off-diagonal component of the optical tensor
(and hence an opposite Farady/Kerr rotation)
at the energies $\Delta_{\rm C}$, $\Delta_{\rm D}$
compared with the one predicted at the A-exciton energy scale.
The absolute intensity of these additional features in the off-diagonal component of the optical tensor
is found:
\begin{eqnarray}
I_{\rm C}
&=&
\sigma_0
\frac{v_{\rm T/B}^2}{\hbar c_{\rm C}\Delta_{\rm C}}
E_{\rm P,C}
=
4\pi a^2 
\sigma_0
\frac{v_{\rm T/B}^2}{\hbar \Delta_{\rm C}}
 n_{\rm pump}
,
\label{IkerrC}
\\
I_{\rm D}
&=&
\sigma_0
\frac{v_{\rm 0/T}^2}{\hbar c_{\rm D}\Delta_{\rm D}}
E_{\rm P,D}
=
4\pi a^2 
\sigma_0
\frac{v_{\rm 0/T}^2}{\hbar \Delta_{\rm D}}
 n_{\rm pump}
 .
\label{IkerrD}
\end{eqnarray}
The expression of Eqs (\ref{IkerrC}), (\ref{IkerrD}) is formally identical
at Eq. (\ref{IkerrA}) for $I_{\rm A}$,  upon changing the proper variables,
with the noticable difference of a factor 2.
This is due to the fact that the strength of $\sigma_{xy,{\rm pump}}(\omega)$
at $\omega \approx \Delta_{\rm A}$ profits of the presence
of the pump-driven charge in {\em both} the conduction and valence bands.
On the other hand, the onset of a finite off-diagonal component $\sigma_{xy,{\rm pump}}(\omega)$
at the energies $\omega \approx \Delta_{\rm C}$, $\omega \approx \Delta_{\rm D}$
is related in an independent way only to the pump-driven charge
in the conduction band and in the valence band, respectively.
This complex multi-peak structure of the pump-induced Faraday/Kerr effect
opens interesting perspective not only for characterizing and proving
the onset of these effects, but also for harvesting them
for multi-frequency operative devices.

\section{Time-dependence}

In the previous Section
we have shown, using an appropriate three-band model, how
a valley-selective population driven by a circularly-polarized pump
can give rise to an off-diagonal component
of the optical tensor, and hence to a finite Faraday/Kerr optical rotation
at {\em three} characteristic energies, related to
the A-exciton, the C-exciton, and to another energy scale
governed by the optical transitions between the lowest conduction band
and high-energy conduction band, roughly determined by the energy difference
between the A and C-exciton.

Most notable, in this description, is the absence of any Faraday/Kerr signature
at the B-exciton energy.
This is essentially due to the strong light-polarization/orbital/valley/spin
entanglemet, so that a circularly-polarized pumping tuned at the A-exciton resonance
induces valley/spin selective population.
More in particular, a left-circularly polarized light tuned at the A-exciton resonance,
in the absence of scattering, triggers particle-hole transitions
uniquely a the K valley and uniquely for spin-up electrons,
making thus an optical unbalance only in the spin-up sector.
This scenario gives rise to a finite Faraday/Kerr signature
only close to the energies 
$\omega \approx \Delta_{\rm A}$,
$\omega \approx \Delta_{\rm C}$, $\omega \approx \Delta_{\rm D}$.
Such snapshot is valid however only on a short time-scale, before
impurity and many-body scattering can cause spin-flip and/or intervalley processes.

In order to gain an insight about how these many-body scattering can
affect the magneto-optical Faraday/Kerr properties induced
by a circularly-polarized pumping, we consider the charge density
in each relevant band which is involved in the time-dynamics.
We denote thus $n_{3z^2-r^2,\sigma}(\nu)$ the charge density
in the lowest-energy conduction band with $d_{3z^2-r^2}$-orbital
and $\sigma$-spin character at the $\nu$ valley,
$n_{{\rm R},\uparrow}({\rm K})$
the charge (hole) density in the valence band at the K valley
with spin-up, and $n_{{\rm L},\downarrow}({\rm K}^\prime)$
the charge (hole) density in the valence band at the K$^\prime$ valley
with spin-down.
Neglecting the frequency-resolved details of each optical feature,
we can estimate the ``Faraday/Kerr'' intensity of each spectral feature as:
\begin{eqnarray}
I_{\rm A}(t)
&\approx &
n_{3z^2-r^2,\uparrow}({\rm K},t)
+
n_{{\rm R},\uparrow}({\rm K},t)
-
n_{3z^2-r^2,\downarrow}({\rm K}^\prime,t)
-
n_{{\rm L},\downarrow}({\rm K}^\prime,t)
,
\\
I_{\rm C}(t)
&\approx &
n_{{\rm R},\uparrow}({\rm K},t)
-
n_{{\rm L},\downarrow}({\rm K}^\prime,t)
,
\\
I_{\rm D}(t)
&\approx &
n_{3z^2-r^2,\uparrow}({\rm K},t)
-
n_{3z^2-r^2,\downarrow}({\rm K}^\prime,t)
,
\\
I_{\rm B}(t)
&\approx &
n_{3z^2-r^2,\downarrow}({\rm K},t)
-
n_{3z^2-r^2,\uparrow}({\rm K}^\prime,t)
.
\end{eqnarray}
Here following the analysis for the other features, we have properly estimated
the intensity of a spectral feature at the energy corresponding to the B-exciton
as resulting by the $d_{{\rm R},\downarrow}({\rm K}) \leftrightarrow d_{3z^2-r^2,\downarrow}({\rm K})$
and $d_{{\rm L},\uparrow}({\rm K}^\prime) \leftrightarrow d_{3z^2-r^2,\uparrow}({\rm K}^\prime)$,
and hence governed by the time-dynamics of 
$n_{3z^2-r^2,\downarrow}({\rm K})$ and $n_{3z^2-r^2,\uparrow}({\rm K}^\prime)$.
In all the cases, we have taken into account that,
due to the orbital/spin/valley entanglement,
similar processes at K$^\prime$ cancel the contributions at the K valley.
Assuming an initial pumping with left-circularly polarized photons resonant
at the A-exciton energy,
we model at $t=0$ the respective charge density as:
$n_{3z^2-r^2,\uparrow}({\rm K},0)=n_{{\rm R},\uparrow}({\rm K},0)=n_{\rm pump}$,
$n_{3z^2-r^2,\downarrow}({\rm K},0)=n_{3z^2-r^2,\downarrow}({\rm K}^\prime,0)=
n_{3z^2-r^2,\uparrow}({\rm K}^\prime,0)=n_{{\rm L},\downarrow}({\rm K}^\prime,0)
=0$.
These conditions reproduce the static results discussed in the previous Section.

\begin{figure}[t!]
\begin{center}
\includegraphics[width=0.95\textwidth]{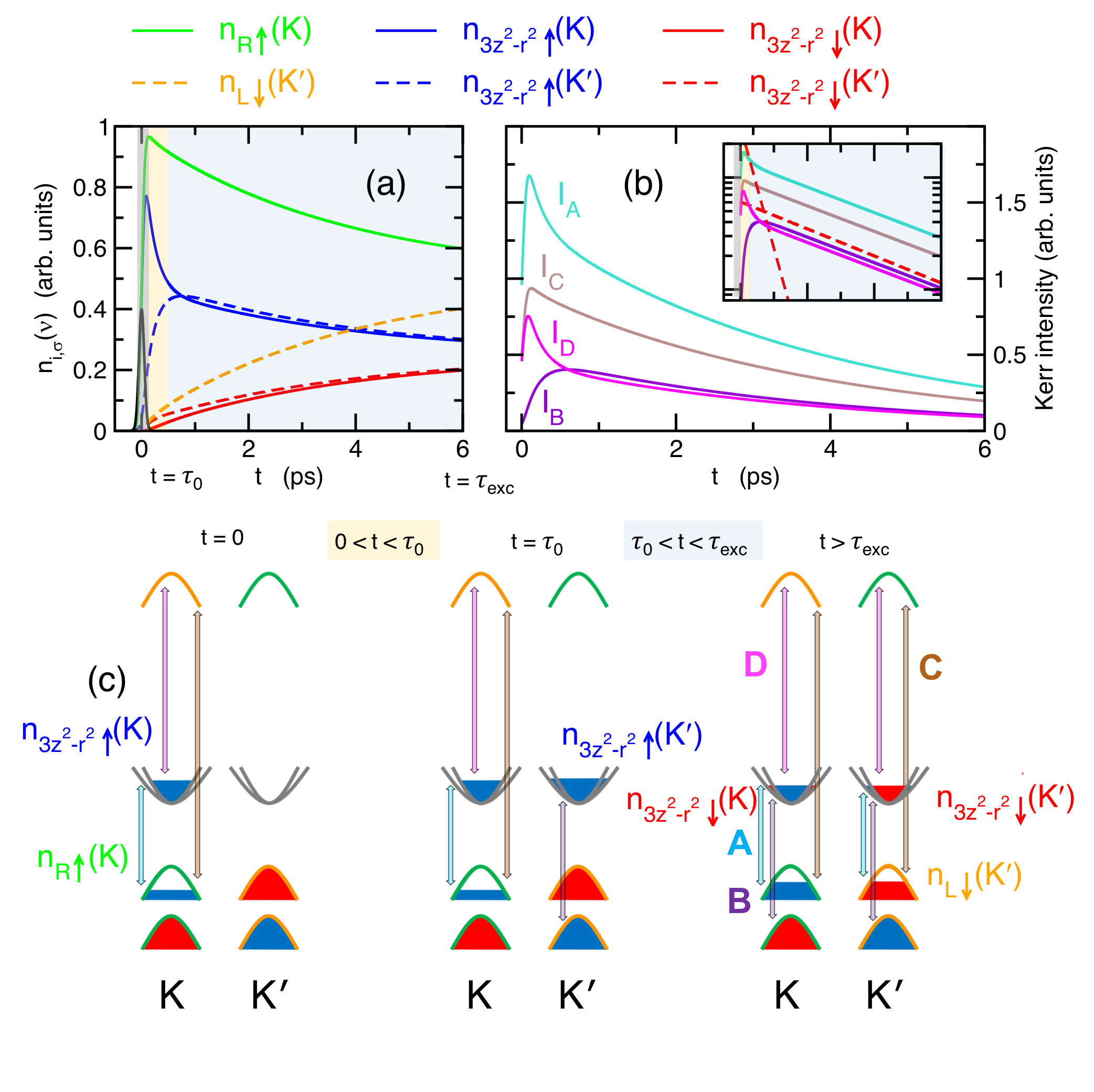}
\end{center}
\caption{({\bf a}) Time-dependence of the pump-driven charge densities $n_{i,\sigma}(\nu)$
upon a Gaussian pump with width $\sim 50$ fs (black solid line).
Legend for different charge densities is reported above.
({\bf b}) Corresponding time-dependence of the integrated Faraday/Kerr intensities
of the different spectral features associated
with different optical interband transitions.
Inset: the same on a linear-log plot, where the red dashed lines
represent the exponential behaviors  decay rate $2\tau_0$ and $2\tau_{\rm exc}$ respectively.
({\bf c}) Sketch of the spin-resolved charge densities $n_{i,\sigma}(\nu)$
in the conduction and valence bands at the K and K$^\prime$ points
along with the main optical transitions
in the different time regimes as outlined in the panels above.
}
\label{f-time}
\end{figure}

Recombination processes, related to annihilation of particle-hole excitations,
are  known to occur on a very long time scale.
On the other hand, the off-diagonal term $\sigma_{xy,{\rm pump}}(\omega)$ of the optical tensor 
is expected to vanish on a much shorter time scale when
scattering processes redistribute the charge in both the conduction and valence bands
giving identical populations in the K, K$^\prime$ valley.
Neglecting the very weak intravalley spin-flip processes,
two main scattering channels have been identified in this scenario \cite{mak12nn,Mai_2014,yang15,Bertoni_2016,schmidt16,molina17,Yan_2017}:
($i$) an interband spin-conserving scattering,
mediated by electron-phonon coupling and/or impurities,
where the charge-density of the conduction band with given spin
is scattered for a valley $\nu$ to the opposite valley $-\nu$.
This process is hampered in the valence band due to the
spin-splitting \cite{Mai_2014,molina17,Yan_2017}; ($ii$) spin-flip intervalley exchange where a particle-hole couple
in a given valley with given spin is scattered into the opposite valley
with reverse spin \cite{yu2014,molina17,schmidt16}.
According with this picture,
we model in a compact way the time dynamics of the pump-driven charges
with a set of coupled equations:
\begin{eqnarray}
\frac{dn_{3z^2-r^2,\uparrow}({\rm K})}{dt}
&=&
\alpha f(t)
-\frac{\mbox{min}[n_{3z^2-r^2,\uparrow}({\rm K}),n_{{\rm R},\uparrow}({\rm K})]
-
\mbox{min}[n_{3z^2-r^2,\downarrow}({\rm K}^\prime),n_{{\rm L},\downarrow}({\rm K}^\prime)]
}{\tau_{\rm exc}}
\nonumber\\
&&
-\frac{n_{3z^2-r^2,\uparrow}({\rm K})-n_{3z^2-r^2,\uparrow}({\rm K}^\prime)
}{\tau_{\rm 0}}
,
\label{et1}
\\
\frac{dn_{3z^2-r^2,\downarrow}({\rm K})}{dt}
&=&
-
\frac{n_{3z^2-r^2,\downarrow}({\rm K})-n_{3z^2-r^2,\downarrow}({\rm K}^\prime)
}{\tau_{\rm 0}}
,
\\
\frac{dn_{{\rm R},\uparrow}({\rm K})}{dt}
&=&
\alpha f(t)
-\frac{\mbox{min}[n_{3z^2-r^2,\uparrow}({\rm K}),n_{{\rm R},\uparrow}({\rm K})]
-
\mbox{min}[n_{3z^2-r^2,\downarrow}({\rm K}^\prime),n_{{\rm L},\downarrow}({\rm K}^\prime)]
}{\tau_{\rm exc}}
,
\\
\frac{dn_{3z^2-r^2,\uparrow}({\rm K}^\prime)}{dt}
&=&
-\frac{n_{3z^2-r^2,\uparrow}({\rm K}^\prime)-n_{3z^2-r^2,\uparrow}({\rm K})
}{\tau_{\rm 0}}
,
\\
\frac{dn_{3z^2-r^2,\downarrow}({\rm K}^\prime)}{dt}
&=&
-\frac{\mbox{min}[n_{3z^2-r^2,\downarrow}({\rm K}^\prime),n_{{\rm L},\downarrow}({\rm K}^\prime)]
-\mbox{min}[\mbox{min}[n_{3z^2-r^2,\uparrow}({\rm K}),n_{{\rm R},\uparrow}({\rm K})]}{\tau_{\rm exc}}
\nonumber\\
&&
-\frac{n_{3z^2-r^2,\downarrow}({\rm K}^\prime)-n_{3z^2-r^2,\downarrow}({\rm K})}{\tau_{\rm 0}}
,
\\
\frac{dn_{{\rm L},\downarrow}({\rm K}^\prime)}{dt}
&=&
-\frac{\mbox{min}[n_{3z^2-r^2,\downarrow}({\rm K}^\prime),n_{{\rm L},\downarrow}({\rm K}^\prime)]
-\mbox{min}[\mbox{min}[n_{3z^2-r^2,\uparrow}({\rm K}),n_{{\rm R},\uparrow}({\rm K})]}{\tau_{\rm exc}}
,
\label{et6}
\end{eqnarray}
where $f(t)$ is the profile of the pump pulse, $\alpha$ is related to the absorption coefficient,
and $\tau_{\rm exc}$, $\tau_{\rm 0}$ are the scattering rates of the two processes discussed above.
The factors $\mbox{min}[n_{3z^2-r^2,\uparrow}({\rm K}),n_{{\rm R},\uparrow}({\rm K})]$
and  $\mbox{min}[n_{3z^2-r^2,\downarrow}({\rm K}^\prime),n_{{\rm L},\downarrow}({\rm K}^\prime)]$
take into account
that the intervalley exchange scattering requires the presence of both particle-hole changes
in the conduction and valence bands.
We take in the following representative values
$\tau_{\rm 0}=200$ fs \cite{yang15,bertoni16}
and $\tau_{\rm exc}=1.8$ ps \cite{yu2014,zhu14}.
The time dynamics of the different charge densities $n_{i,\sigma}(\nu)$
is shown in Fig. \ref{f-time}a, and the corresponding time-dependence of the Kerr intensity
of the different spectral features in Fig. \ref{f-time}b, whereas panel (c)
depicts some representative time-snapshots of $n_{i,\sigma}(\nu)$.
Notice that, in order to focus on the time dynamics,
we plot here only the dependence of $I_\mu(t)$ on the time-dependent charge densities $n_{i,\sigma}(\nu,t)$,
neglecting the current operator matrix elements and other time-independent factors,
so that the relative ratio of the intensities here is not meant to be representative
of the experimental ratio.
The overall behavior of $n_{i,\sigma}(\nu)$ and
$I_{\rm A}(t)$ that we get is in very good agreement with the results
by Ref. \cite{molina17}, performed with {\em ab-initio}
techniques.

We can identify there three main regimes:
($i$) a short-time scale $t \ll \tau_0$ (grey areas, left panel of Fig. \ref{f-time}c) where the charge populations
are mainly determined by the driving pump,
with a significant population of only $n_{3z^2-r^2,\uparrow}({\rm K})$ and $n_{{\rm R},\uparrow}({\rm K})$.
This is reflected in a sharp onset of the Farady/Kerr intensities
$I_{\rm A}(t)$, $I_{\rm C}(t)$ and $I_{\rm D}(t)$.
($ii$)
soon after this scenario, in the short time-range $t \sim \tau_0 $ (yellow areas, middle panel of Fig. \ref{f-time}c),
the impurity/electron-phonon scattering has a main effect of a depletion
of $n_{3z^2-r^2,\uparrow}({\rm K})$ due a redistribution of the spin-up
conduction electrons towards $n_{3z^2-r^2,\uparrow}({\rm K}^\prime)$.
As a consequence, a sharp decrease of $I_{\rm A}(t) \propto \mbox{e}^{-t/2\tau_0}$ and $I_{\rm D}(t)\propto \mbox{e}^{-t/2\tau_0}$
is predicted, whereas the finite valley population
$n_{3z^2-r^2,\uparrow}({\rm K}^\prime)$ gives rise to a finite (delayed)
intensity of $I_{\rm B}(t)$.
($iii$)
in a following time regime $\tau_0 \ll t \sim \tau_{\rm exc}$ (light blue areas, right panel of Fig. \ref{f-time}c)
the key many-body process is the exchange scattering (assisted by the impurity/electron-phonon one)
leading towards a slower equal spin population of each conduction and valence band.
The total spectral intensities in this regime scale as $I_\mu(t) \propto \mbox{e}^{-t/2\tau_{\rm exc}}$.
The final steady state (right panel of Fig. \ref{f-time}c) is however reached only for $t \gg \tau_{\rm exc}$.
In this regime the contributions of off-diagonal elements of the optical tensor at K and K$^\prime$ cancel out,
and any Faraday/Kerr effect vanishes.
Note that the transient Faraday/Kerr intensity at the B-exciton energy range is a by-product
of the finite valley-K$^\prime$ population of $n_{3z^2-r^2,\uparrow}$.
In similar way as this valley-transient population is expect to give rise
to a Faraday/Kerr effect at the energy $\Delta_{\rm B}$ associated with optical transitions between
$n_{{\rm L},\uparrow}({\rm K}^\prime)$ and $n_{3z^2-r^2,\uparrow}({\rm K}^\prime)$,
we can expect the appearence of further Faraday/Kerr features at
the energies associated with optical transitions between
$n_{{\rm L},\uparrow}({\rm K}^\prime)$ and $n_{{\rm R},\uparrow}({\rm K}^\prime)$
and
between $n_{3z^2-r^2,\uparrow}({\rm K}^\prime)$ and $n_{{\rm R},\uparrow}({\rm K}^\prime)$.
We denote these transitions as $\Delta_{\rm C}^\prime$
and $\Delta_{\rm D}^\prime$, whose Faraday/Kerr spectral intensity
scales, assuming pumping tuned at the A-exciton resonance, as 
$I_{\rm C}^\prime(t) \propto I_{\rm C}(t)$
and $I_{\rm D}^\prime(t) \propto I_{\rm D}(t)$.
The band-parameters determining the detailed spectral properties of these features
are also listed in Table \ref{t:par}.

\section{Conclusions}

In summary, in the present paper we have addressed in an analytical way the onset of a finite Faraday/Kerr effect
in single-layer transition-metal dichalcogenides upon pumping with a circularly polarized light at the A-resonance.
To this aim we have introduced a proper three-band analytical ${\bf k} \cdot {\bf p}$ model
that retains all the orbital complexity of original band-structure and of the optical selection rules.
We have shown how a pump-driven spin/valley selective population gives rise
to a finite off-diagonal component of the optical tensor responsible for different spectral features in the Faraday/Kerr optical rotation.
We recover the signature of a Faraday/Kerr rotation in the proximity of pump energy at the A-exciton resonance,
in accordance with the available experimental and theoretical findings \cite{plechinger14,zhu14,dalconte15,sun16,plechinger17,mccormick18,catarina20},
and we predict the onset of a Faraday/Kerr signal at
the C-exciton resonance and at a further energy scale related to the high-energy conduction band.
These predictions can guide future experimental investigation,
spanning also the different families of $MX_2$ TMDs.
We have further modelled the effects of time-dynamics driven by the many-body scattering,
and the consequent emerging of additional transient Faraday/Kerr optical features.
Our results provide a suitable compact modelling
for describing the magneto-optical properties induced in transition-metal dichalcogenides
by circularly polarized pumping, in terms of few simple intuitive representative parameters.

\vspace{6pt} 





\funding{
E.C. acknowledges financial support from PNRR MUR Project No. PE0000023-NQSTI.
}



\acknowledgments{\mbox{}}

\conflictsofinterest{The authors declare no conflicts of interest.} 



\appendixtitles{no} 
\appendixstart
\appendix

\section{Mapping ${\bf k}\cdot {\bf p}$ band parameters
in terms of TB parameters}

\label{a:TB}

Below we summarized useful analytical expressions for the band-parameters
of the ${\bf k}\cdot {\bf p}$ Hamiltonian in Eqs. (\ref{HpK})-(\ref{HpKp})
in terms of the original tight-binding parameters provided in Ref. \cite{3bands}.

We have explicitely:
\begin{eqnarray}
E_{\rm 0}
&=&
\epsilon_1
-3(t_0-2r_0+u_0)
,
\\
E_{\rm T}
&=&
\epsilon_2
-\frac{3}{2}(t_{11}+t_{22}-4r_{11}+u_{11}+u_{22})
+2\sqrt{3}r_{12}
+
3\sqrt{3}t_{12}
-
3\sqrt{3}u_{12}
,
\\
E_{\rm B}
&=&
\epsilon_2
-\frac{3}{2}(t_{11}+t_{22}-4r_{11}+u_{11}+u_{22})
+2\sqrt{3}r_{12}
-
3\sqrt{3}t_{12}
+
3\sqrt{3}u_{12}
,
\\
a_{0}
&=&
\frac{3(t_0-6r_0+4u_0)}{4}
,
\\
a_{\rm T}
&=&
\frac{3}{8}(t_{11}+4u_{11}+t_{22}+4u_{22})
-\frac{9}{2}r_{11}
-\frac{3\sqrt{3}}{2}r_{12}
-
\frac{3\sqrt{3}}{4}t_{12}
+
3\sqrt{3}u_{12}
,
\\
a_{\rm B}
&=&
\frac{3}{8}(t_{11}+4u_{11}+t_{22}+4u_{22})
-\frac{9}{2}r_{11}
-\frac{3\sqrt{3}}{2}r_{12}
+
\frac{3\sqrt{3}}{4}t_{12}
-
3\sqrt{3}u_{12},
\\
v_{\rm 0/T}
&=&
-\frac{3\sqrt{3}}{2\sqrt{2}}
t_2
+
\frac{3\sqrt{3}}{\sqrt{2}}
u_2
+
\frac{3}{2\sqrt{2}}
t_1
-
\frac{3(r_1-r_2)}{\sqrt{2}}
+
\frac{3}{\sqrt{2}}u_1
,
\\
v_{\rm 0/B}
&=&
-\frac{3\sqrt{3}}{2\sqrt{2}}
t_2
+
\frac{3\sqrt{3}}{\sqrt{2}}
u_2
-
\frac{3}{2\sqrt{2}}
t_1
+
\frac{3(r_1-r_2)}{\sqrt{2}}
-
\frac{3}{\sqrt{2}}u_1
,
\\
v_{\rm T/B}
&=&
\frac{3\sqrt{3}}{4}
(t_{11}-t_{22}-2u_{11}+2u_{22})
.
\end{eqnarray}

The terms $\bar{a}_{{\rm T},\sigma}$, $\bar{a}_{0,\sigma}$, $\bar{a}_{{\rm B},\sigma}$
are related to the dispersion mass of each band, and they are obtained
within the ${\bf k}\cdot {\bf p}$ context as:
\begin{eqnarray}
\bar{a}_{{\rm T},\uparrow}
&=&
a_{\rm T}p^2
+\gamma_{{\rm T/B},\uparrow} 
+\gamma_{{\rm T/0},\uparrow} 
,
\\
\bar{a}_{{\rm 0},\uparrow}
&=&
a_{0}
+\gamma_{{\rm 0/B},\uparrow} 
-\gamma_{{\rm T/0},\uparrow}
,
\\
\bar{a}_{{\rm B},\uparrow}
&=&
a_{\rm B}p
-\gamma_{{\rm T/B},\uparrow} 
-\gamma_{{\rm 0/B},\uparrow} 
,
\\
\bar{a}_{{\rm T},\downarrow}
&=&
a_{\rm T}p^2
+\gamma_{{\rm T/B},\downarrow} 
+\gamma_{{\rm T/0},\downarrow} 
,
\\
\bar{a}_{{\rm 0},\downarrow}
&=&
a_{0}
+\gamma_{{\rm 0/B},\downarrow} 
-\gamma_{{\rm T/0},\downarrow}
,
\\
\bar{a}_{{\rm B},\downarrow}
&=&
a_{\rm B}p
-\gamma_{{\rm T/B},\downarrow} 
-\gamma_{{\rm 0/B},\downarrow} 
,
\end{eqnarray}
and where
\begin{eqnarray}
\gamma_{{\rm T/B},\sigma}
&=&
\frac{v_{T/B}^2}{E_{\rm T}-E_{\rm B}+2\lambda I_\sigma}
,
\\
\gamma_{{\rm T/0},\sigma} 
&=&
\frac{v_{0/T}^2}{E_{\rm T}-E_0 + \lambda I_\sigma}
,
\\
\gamma_{{\rm 0/B},\sigma} 
&=&
\frac{v_{0/B}^2}{E_0-E_{\rm B} + \lambda I_\sigma}
.
\end{eqnarray}
The spin-dependent single-particle energies
at the K point read thus:
\begin{eqnarray}
\epsilon_{T,\uparrow}
&=&
E_{\rm T}-\lambda
,
\\
\epsilon_{0,\uparrow}
&=&
E_{0}
,
\\
\epsilon_{{\rm B},\uparrow}
&=&
E_{\rm B}+\lambda
,
\\
\epsilon_{T,\downarrow}
&=&
E_{\rm T}+\lambda
,
\\
\epsilon_{0,\downarrow}
&=&
E_{0}
,
\\
\epsilon_{{\rm B},\downarrow}
&=&
E_{\rm B}-\lambda
,
\end{eqnarray}
and the characteristic
interband optical edges:
\begin{eqnarray}
\Delta_{\rm A}
&=&
\epsilon_{0,\uparrow}
-
\epsilon_{{\rm B},\uparrow}
,
\\
\Delta_{\rm B}
&=&
\epsilon_{0,\downarrow}
-
\epsilon_{{\rm B},\downarrow}
,
\\
\Delta_{\rm C}
&=&
\epsilon_{{\rm T},\uparrow}
-
\epsilon_{{\rm B},\uparrow}
,
\\
\Delta_{\rm D}
&=&
\epsilon_{{\rm T},\uparrow}
-
\epsilon_{{\rm 0},\uparrow}
,
\\
\Delta_{\rm C^\prime}
&=&
\epsilon_{{\rm T},\downarrow}
-
\epsilon_{{\rm B},\downarrow}
,
\\
\Delta_{\rm D^\prime}
&=&
\epsilon_{{\rm T},\downarrow}
-
\epsilon_{{\rm 0},\downarrow}
.
\end{eqnarray}

In Table \ref{t:par} we summarize the numerical values
of our three-band ${\bf k}\cdot {\bf p}$ model
obtained from the initial tight-binding parameters of Ref. \cite{3bands}:
\begin{table}[t] 
\caption{This is a table caption.
\label{t:par}
}
\newcolumntype{c}{>{\centering\arraybackslash}X}
\begin{tabularx}{\textwidth}{c|cccccc}
\toprule
 & \bf MoS$_2$ & \bf MoSe$_2$ & \bf MoTe$_2$ & \bf WS$_2$ & \bf WSe$_2$ & \bf WTe$_2$ \\
\midrule
$\lambda$ & 0.073 & 0.091 & 0.107 & 0.211 & 0.228 & 0.237  \\
$E_{\rm T}$ & 3.451 & 3.056 & 2.525 & 3.933 & 3.443 &  2.871 \\
$E_{0}$ & 1.595 & 1.482 & 1.113 & 1.749 & 1.565 &  1.132 \\
$E_{\rm B}$ & -0.062 & 0.053 & 0.041 & -0.057 & 0.024 &  0.065 \\
$\epsilon_{{\rm T},\uparrow}$ & 3.378 & 2.965 & 2.418 & 3.722 & 3.215 & 2.634 \\
$\epsilon_{0,\uparrow}$ & 1.595 & 1.482 & 1.113 & 1.749 & 1.565 & 1.132 \\
$\epsilon_{{\rm B},\uparrow}$ & 0.011 & 0.144 & 0.148 & 0.154 & 0.252 & 0.302 \\
$\epsilon_{{\rm T},\downarrow}$ & 3.524 & 3.147 & 2.632 & 4.144 & 3.671 & 3.108 \\
$\epsilon_{0,\downarrow}$ & 1.595  & 1.482 & 1.113 & 1.749 & 1.565 &  1.132 \\
$\epsilon_{{\rm B},\downarrow}$ & -0.135 & -0.038 & -0.066 & -0.268 & -0.204 & -0.172  \\
$\Delta_{\rm A}$ & 1.584 & 1.338 & 0.965 & 1.595 & 1.313 & 0.830 \\
$\Delta_{\rm B}$ & 1.730 & 1.520 & 1.179 & 2.017 & 1.769 & 1.304 \\
$\Delta_{\rm C}$ & 3.367 & 2.821 & 2.270 & 3.569 & 2.963 & 2.332 \\
$\Delta_{\rm D}$ & 1.783 & 1.483 & 1.305 & 1.973 & 1.650 & 1.502\\
$\Delta_{\rm C^\prime}$ & 3.659 & 3.185 & 2.698 & 4.413 & 3.875 & 3.280 \\
$\Delta_{\rm D^\prime}$ & 1.929 & 1.665 & 1.519 & 2.395 & 2.106 & 1.976 \\
$a_{\rm T}$ & -1.190 & -1.170 & -1.142 & -1.305 & -1.310 & -1.221 \\
$a_{\rm 0}$ &-0.493 & -0.411 & -0.012 & -0.586 & -0.455 &  -0.212 \\
$a_{\rm B}$ & 1.060 & 0.921 & 0.578 & 1.299 & 1.119 &  0.860 \\
$\bar{a}_{{\rm T},\uparrow}$ & -0.800 & -0.774 & -0.768 & -0.836 & -0.829 & -0.816 \\
$\bar{a}_{0,\uparrow}$ & 0.961 & 0.800 & 0.873 & 1.512 & 1.375 & 1.624 \\
$\bar{a}_{{\rm B},\uparrow}$ & -0.777 & -0.675 & -0.667 & -1.245 & -1.161 & -1.346  \\
$\bar{a}_{{\rm T},\downarrow}$ & -0.831 & -0.819 & -0.826 & -0.925 & -0.941 &  -0.932 \\
$\bar{a}_{0,\downarrow}$ & 0.838 & 0.654 & 0.710 & 1.071 & 0.902 &  0.953 \\
$\bar{a}_{{\rm B},\downarrow}$ & -0.637 & -0.505 & -0.470 & -0.756 & -0.628 & -0.615  \\
$v_{0/{\rm B}}$ & -1.545 & -1.292 & 0.956 & -1.850 & -1.565 &  -1.244 \\
$v_{0/{\rm T}}$ & -0.306 & -0.236 & 0.286 & -0.303 & -0.246 &  -0.204 \\
$v_{{\rm T/B}}$ & -1.065 & -1.001 & 0.841 & -1.228 & -1.148 &  -0.938 \\
$c_{\rm A}$ & 1.739 & 1.475 & 1.540 & 2.757 & 2.535 & 2.969 \\
$c_{\rm B}$ & 1.474 &1.160 & 1.180 & 1.828 & 1.530 & 1.568 \\
$c_{\rm C}$ & -0.023 & -0.098 & -0.101 & 0.409 & 0.332 & 0.530 \\
$c_{\rm D}$ & -1.762 & -1.573 & -1.641 & -2.348 & -2.204 & -2.439 \\
$c_{\rm C^\prime}$ & -0.194 & -0.314 & -0.356 & -0.168 & -0.313 & -0.317 \\
$c_{\rm D^\prime}$ & -1.669 & -1.473 & -1.536 & -1.996 & -1.843 & -1.884 \\
\bottomrule
\end{tabularx}
\end{table}

\begin{adjustwidth}{-\extralength}{0cm}

\reftitle{References}

\bibliography{refs}

\PublishersNote{}
\end{adjustwidth}
\end{document}